# Equations of state of new boron-rich selenides $B_6Se$ and $B_{12}Se$

Kirill A. Cherednichenko,[1,2] Yann Le Godec,[3] and Vladimir L. Solozhenko [1,*]

[1] *LSPM–CNRS, Université Sorbonne Paris Nord, 93430 Villetaneuse, France*
[2] *Department of Physical and Colloid Chemistry, Gubkin University, Moscow, 119991, Russia*
[3] *IMPMC–CNRS, UPMC Sorbonne Universités, 75005 Paris, France*

**Abstract**

*Two novel of boron-rich selenides, orthorhombic $B_6Se$ and rhombohedral $B_{12}Se$, have been recently synthesized at high pressure – high temperature conditions. Room-temperature compressibilities of these phases were studied in a diamond anvil cell using synchrotron powder X-ray diffraction. A fit of experimental p-V data by third-order Birch-Murnaghan equation of state yielded the bulk moduli of 155(2) GPa for $B_{12}Se$ and 144(3) GPa for $B_6Se$. No pressure-induced phase transitions have been observed in the studied pressure range, i.e. up to 35 GPa.*

**Introduction**

Boron-rich compounds containing $B_{12}$-icosahedra have been known since 1960-s [1]. The majority of these compounds are the structural derivatives of elemental α-rhombohedral boron (α-$B_{12}$) with general stoichiometry: $B_{12}X_y$ (where X is interstitial atom, $y \leq 2$). They received a considerable attention due to the high hardness, chemical inertness, thermal conductivity, radiation resistance and unusual electronic properties [2-5]. For instance, boron carbide $B_{12}C_3$ is hard compound widely used as abrasive and armor material [6,7], boron suboxide $B_{12}O_2$ is the hardest known oxide [8,9], and $B_{12}As_2$ was found to be extremely stable under electron bombardment and able to the self-healing [10].

Despite a great number of boron-rich compounds discovered during last decades [1,3], their physical properties still remain unknown. Very recently two new boron-rich selenides, rhombohedral $B_{12}Se$ [11] and orthorhombic $B_6Se$ [12], have been synthesized at high pressures and high temperatures. The crystal structures of these compounds have been refined by synchrotron X-ray diffraction studies combined with *ab initio* calculations. The theoretically predicted bulk modulus of $B_6Se$ is 144 GPa [12], however this value has not been experimentally supported so far. Here we report a study of the compressibilities of $B_{12}Se$ and $B_6Se$ in a diamond anvil cell up to 35 GPa in the neon quasi-hydrostatic medium.

---

* Corresponding author: vladimir.solozhenko@univ-paris13.fr

**Experimental**

$B_{12}Se$ and $B_6Se$ were synthesized at 6.1 GPa and 2600-2700 K by reaction of amorphous boron (Grade I ABCR) with elemental selenium (Alfa Aesar, 99.5%)) in a toroid-type high-pressure apparatus. Synthesis details are described elsewhere [11,12].

The study of $B_{12}Se$ and $B_6Se$ compressibilities was performed at Xpress beamline (Elettra) in the membrane diamond anvil cells with 300-μm culet anvils. In order to provide the same compression conditions, the polycrystalline samples of both compounds were mixed and loaded together with small ruby balls into 150-μm hole drilled in a rhenium gasket pre-indented down to 25 μm. Neon was chosen as pressure transmitting medium providing quasi-hydrostatic compression. Pressure was determined *in situ* from the calibrated shift of the ruby R1 fluorescent line [15]; the pressure drift at each pressure point did not exceed 0.6 GPa. The synchrotron radiation was set to a wavelength of 0.4957 Å using channel-cut Si (111) monochromator and focused down to 20 μm. The X-ray diffraction patterns were collected in the 2-23 2θ-range using MAR 345 image plate detector with an exposure time of 600 seconds. The diffraction patterns were further processed using FIT2D [16] and Powder Cell [17] software; the lattice parameters of $B_{12}Se$ and $B_6Se$ at different pressures are presented in Table S1.

**Results and discussion**

The unit cells of both boron-rich selenides are presented in Fig. 1. Rhombohedral $B_{12}Se$ is a typical α-$B_{12}$-related boron-rich compound with $B_{12}$-icosahedra placed in the corners and on one of the main diagonals of the hexagonal unit cell. As was previously found [11], the occupation of *6c* site by Se atom is ~50%, which can be explained by the short Se-Se distance (2.02 Å) compared with the covalent radius of selenium (1.20 Å). $B_6Se$ has an orthorhombic crystal structure containing $B_{12}$-icosahedra with side-centered packing in the unit cell similar to that in $B_3Si$ [18]. Unlike $B_{12}Se$, the occupancy of Se atoms (in *4h* Wykoff positions) in $B_6Se$ was found to be close to 100% [12]. The latter resulted in a higher $B_6Se$ density (3.58 g/cm$^3$) compared to that of $B_{12}Se$ (2.90 g/cm$^3$).

During compression the reflections of $B_{12}Se$ and $B_6Se$ monotonously shifted towards larger 2θ-values, and no evidence of any phase transition has been observed for both phases over the whole studied pressure range (Fig. 2).

We employed one-dimensional analogue of the first-order Murnaghan equation of state [19] in order to approximate of the nonlinear relation between relative lattice parameters of $B_{12}Se$ and $B_6Se$ unit cells and pressure (Fig. 3).

$$r = r_0 \left[1 + P \left(\frac{\beta'_{0,r}}{\beta_{0,r}}\right)\right]^{-\frac{1}{\beta'_{0,r}}} \quad (1),$$

where $r$ is the lattice parameter (index 0 refers to ambient pressure), $\beta_{0,r}$ is the axial modulus and $\beta'_{0,r}$ is its pressure derivative. The axis moduli and the corresponding pressure derivatives that best fit the experimental data are collected in Table I. According to these data, the compression of $B_{12}Se$ unit cell is quasi isotropic (similar to that of $B_{12}As_2$ [20]), whereas compression of $B_6Se$ is noticeably anisotropic with the highest compressibility along $b$-axis.

The changes of $B_{12}Se$ and $B_6Se$ unit cell volumes under pressure are shown in Fig. 4. The third-order Birch-Murnaghan [21] equation of state was used to fit the experimental data:

$$P(V) = \frac{3B_0}{2}\left[\left(\frac{V_0}{V}\right)^{\frac{7}{3}} - \left(\frac{V_0}{V}\right)^{\frac{5}{3}}\right]\left\{1 + \frac{3}{4}(B'_0 - 4)\left[\left(\frac{V_0}{V}\right)^{\frac{2}{3}} - 1\right]\right\} \qquad (2)$$

The bulk moduli ($B_0$) and corresponding first pressure derivatives ($B_0'$) are 154.6±2.2 GPa and 5.9±0.2 for $B_{12}Se$ and 143.9±2.6 GPa and 4.0±0.2 for $B_6Se$.

The experimentally determined bulk modulus of $B_{12}Se$ was found to be in agreement with the 147 GPa value theoretically predicted in the framework of the thermodynamic model of hardness [22]. According to our findings, $B_{12}Se$ has the lowest bulk modulus value among all boron-rich compounds with structure related to α-rhombohedral boron (Fig. 5). Our results support the general trend – increase of covalent radius of interstitial atom in the row of boron-rich chalcogenides (α-$B_{12}$ [23] – $B_{12}O_2$ [24] – $B_{12}Se$) leads to decrease of bulk modulus – which has been already observed in the row of boron-rich pnictides (α-$B_{12}$ [23] – $B_{13}N_2$ [25] – $B_{12}P_2$ [26] – $B_{12}As_2$ [20]). However, as it follows from Fig. 5, the ways of such decrease for boron-rich chalcogenides and boron-rich pnictides are different (red and black dashed lines, respectively). It is not surprising that the bulk moduli of $B_{12}As_2$ (150.1 GPa) and $B_{12}Se$ (154.6 GPa) containing interstitial atoms with very close covalent radii ($r_{As}$ = 1.19 Å, $r_{Se}$ = 1.20 Å) are very close to each other.

The experimental bulk modulus (144(3) GPa) and X-ray density (3.58 g/cm$^3$) of $B_6Se$ are in good agreement with the theoretically predicted values, i.e. 144 GPa and 3.66 g/cm$^3$ [12]. Despite rather high density, orthorhombic $B_6Se$ remains yet the most compressible boron-rich solid.

**Conclusions**

The compressibilities of two new boron-rich selenides, $B_{12}Se$ and $B_6Se$, synthesized at high pressures and high temperatures were studied up to 35 GPa using synchrotron X-ray powder diffraction in a diamond anvil cell. At room temperature, both selenides were found to be stable in the whole studied pressure range, and no pressure-induced phase transitions were observed.

**Acknowledgements**

The authors thank Dr. V.A. Mukhanov (LSPM-CNRS) for the samples synthesis and Dr. B. Joseph (Elettra) for assistance at Xpress beamline. Synchrotron X-ray diffraction experiments were carried out during beamtime allocated to Proposal 20180186 at Elettra Synchrotron. This work was financially supported by the European Union's Horizon 2020 Research and Innovation Program under Flintstone2020 project (grant agreement No 689279).
**ORCID IDs**

Vladimir L. Solozhenko  https://orcid.org/0000-0002-0881-9761

Kirill A. Cherednichenko  https://orcid.org/0000-0002-1868-8232
4

Table I. Bulk moduli ($B_0$) with first derivatives ($B'_0$) and axial moduli ($\beta_{0,r}$) with first derivatives ($\beta'_{0,r}$) of $B_{12}Se$ and $B_6Se$ obtained from approximations of the experimental data using Birch-Murnaghan and one-dimensional analog of Murnaghan EOSs.

|   | $B_{12}Se$ | $B_6Se$ |
|---|---|---|
| $B_0$ (GPa) | 155(2) | 144(3) |
| $B'_0$ | 5.9(2) | 4.0(2) |
| $\beta_{0,a}$, GPa | 456(6) | 483(20) |
| $\beta'_{0,a}$ | 19(1) | 6(2) |
| $\beta_{0,b}$, GPa | – | 340(12) |
| $\beta'_{0,b}$ | – | 11(1) |
| $\beta_{0,c}$, GPa | 480(9) | 528(43) |
| $\beta'_{0,c}$ | 12(1) | 19(4) |

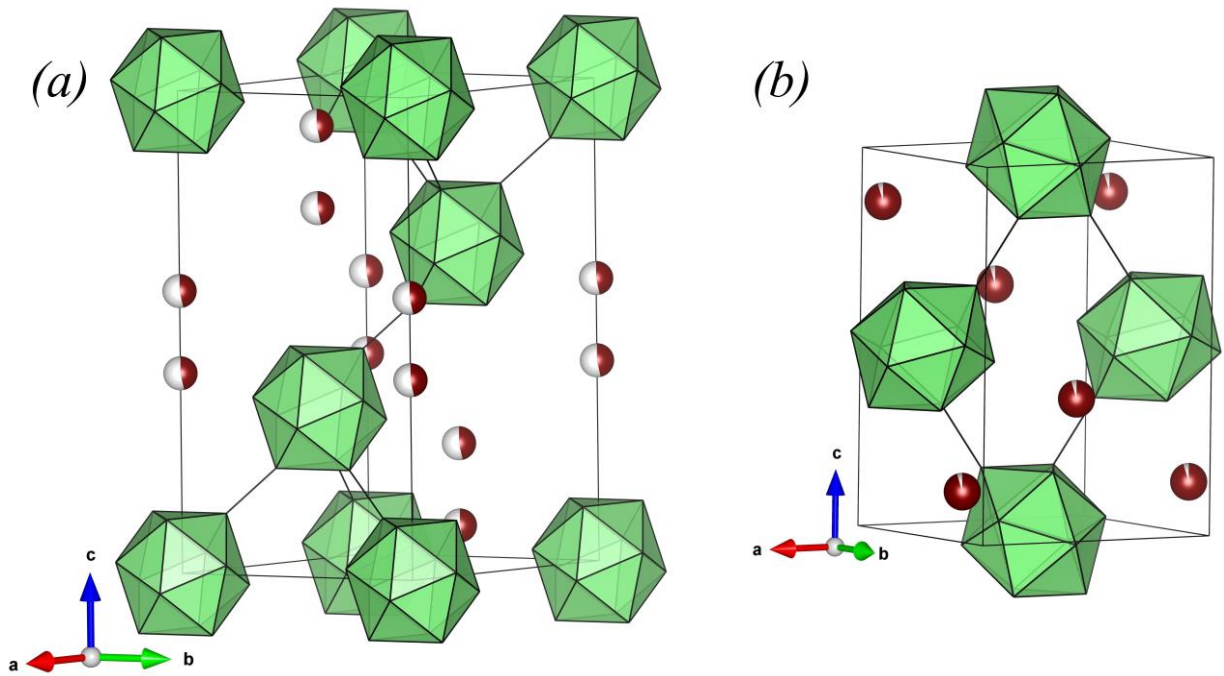

Fig. 1. The unit cells of $B_{12}Se$ (in hexagonal setting) (*a*) and $B_6Se$ (*b*); $B_{12}$-units are presented by green icosahedra, Se atoms are shown as balls.

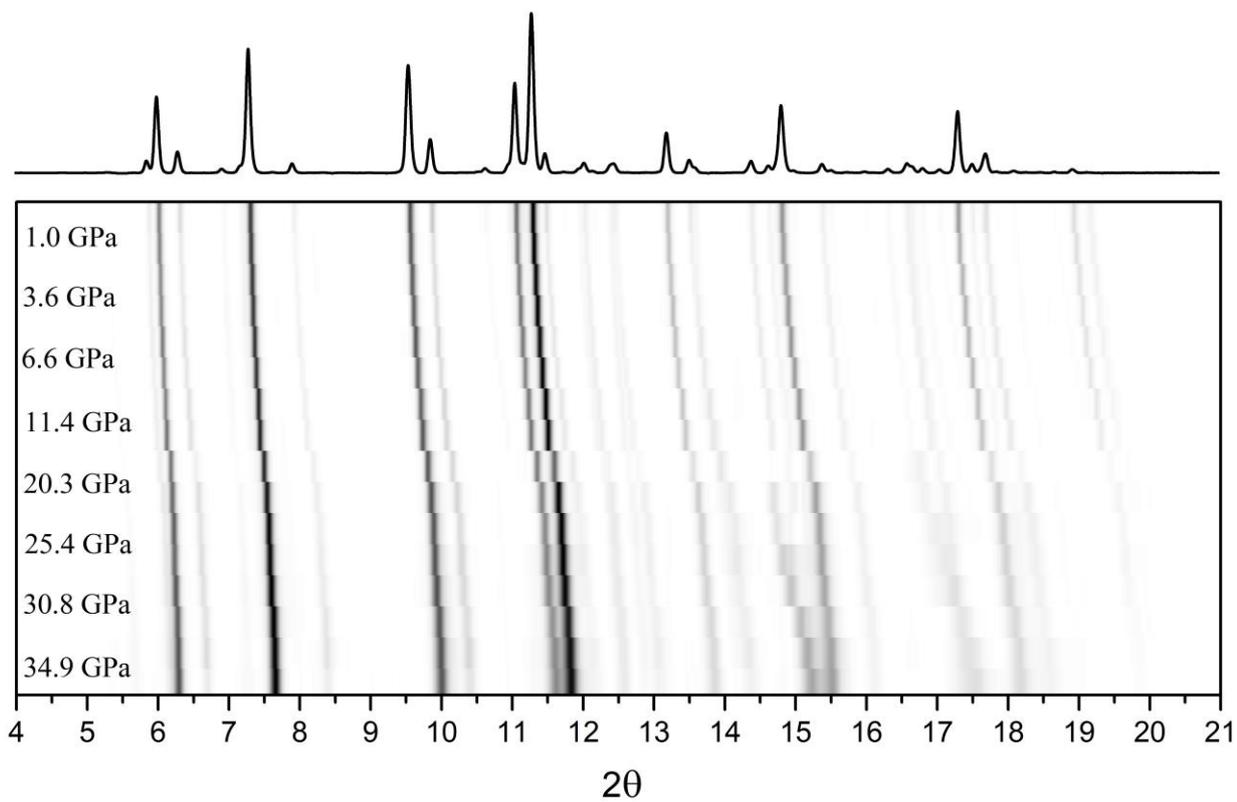

Fig. 2. 2D representation of X-ray diffraction patterns ($\lambda = 0.4957$ Å) of $B_{12}Se/B_6Se$ mixture *versus* pressure.

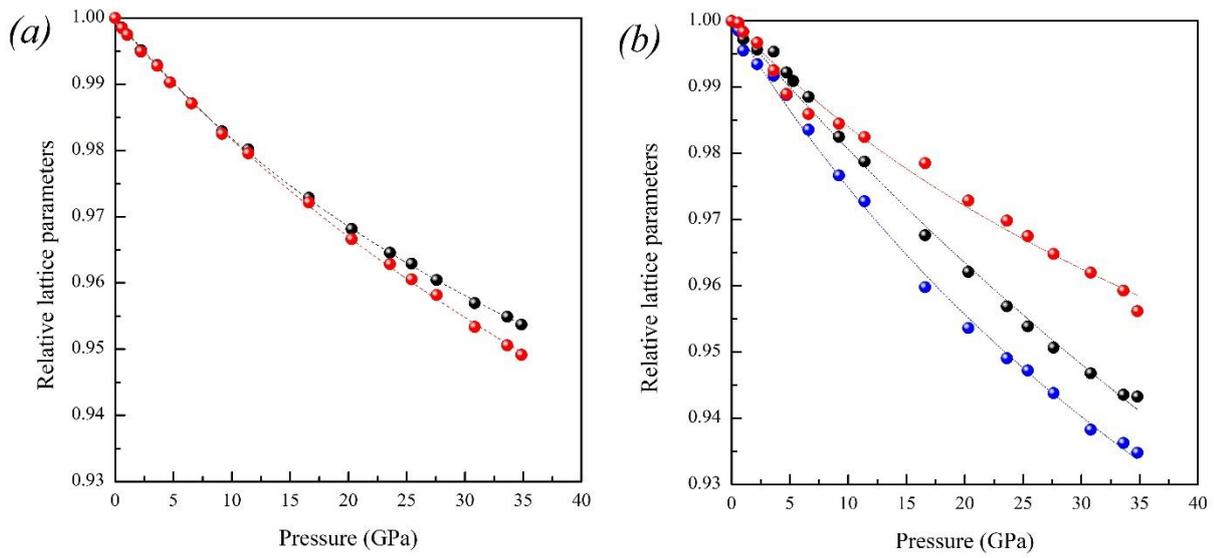

Fig. 3. The relative lattice parameters of $B_{12}Se$ and $B_6Se$ *versus* pressure: $a/a_0$ (black), $b/b_0$ (blue) and $c/c_0$ (red). The dashed lines represent the fits of one-dimensional analog of Murnaghan equation of state to the experimental data.

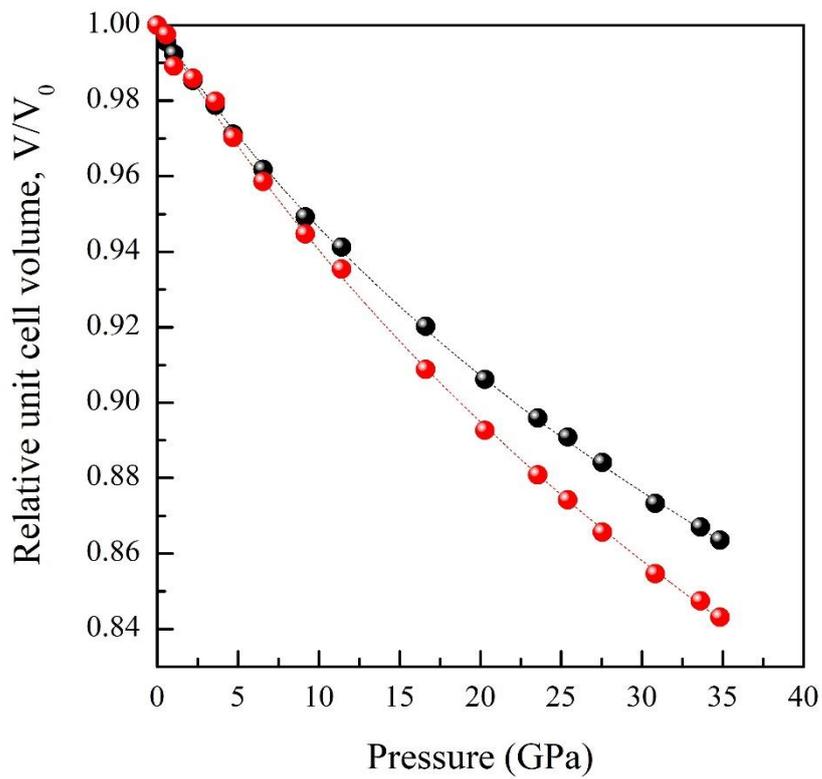

Fig. 4. Equation of state of $B_{12}Se$ (black) and $B_6Se$ (red). The dashed lines represent Birch-Murnaghan equation of state fit to the experimental data.

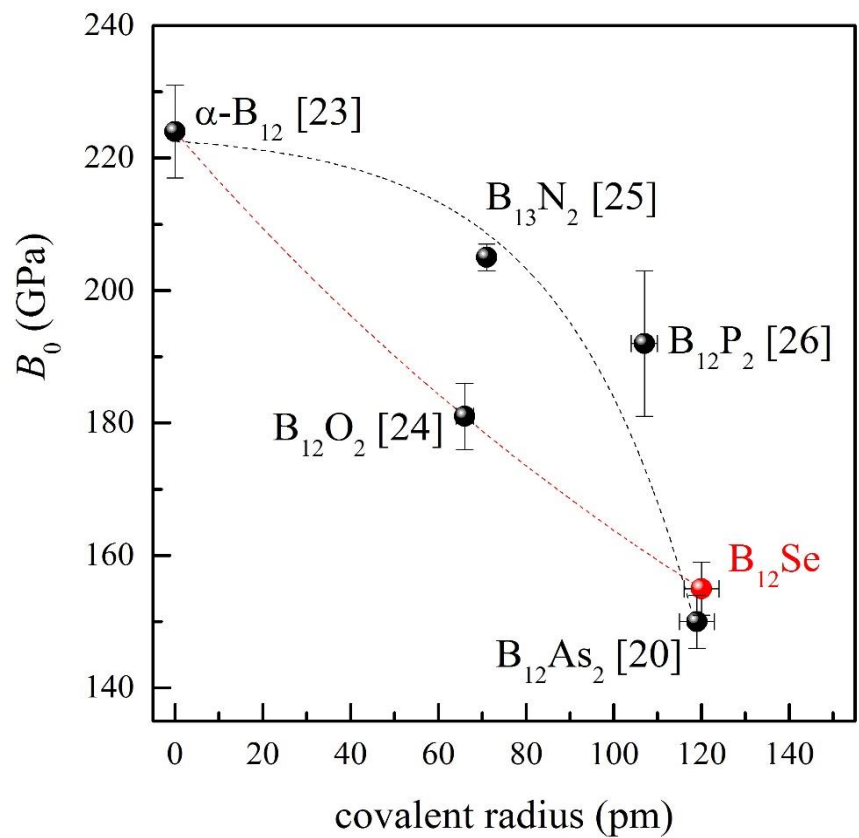

Fig. 5. Bulk moduli of α-rhombohedral boron and related boron-rich compounds.